\documentclass[aps, amsmath, amssymb, preprint, floatfix]{revtex4-1}

\usepackage{graphicx}
\usepackage{color}

\begin{document}

\title{Evolving geometry of a vortex triangle} 
\author{Vikas S. Krishnamurthy}
\affiliation{Departamento de F\'isica, Universidade Federal de Pernambuco, 50670-901 Recife, Brazil}
\email{skvikas@imperial.ac.uk}
\author{Hassan Aref}
\affiliation{Department of Engineering Science \& Mechanics, Virginia Tech, Blacksburg, VA 24061, USA (Deceased)}
\author{Mark A. Stremler}
\affiliation{Department of Biomedical Engineering \& Mechanics, Virginia Tech, Blacksburg, VA 24061, USA}

\date{\today}

\begin{abstract}
The motion of three interacting point vortices in the plane can be thought of as the motion of three geometrical points endowed with a dynamics. This motion can therefore be re-formulated in terms of dynamically evolving geometric quantities, viz.\ the 
circle that circumscribes the vortex triangle and the angles of the vortex triangle. In this study, we develop the equations of motion for the center, $Z$, and radius, $R$, of this circumcircle, and for the angles of the vortex triangle, $A$, $B$, and $C$. The equations of motion for $R$, $A$, $B$ and $C$ form an autonomous dynamical system. A number of known results in the three-vortex problem follow readily from the equations, giving a new geometrical perspective on the problem.
%
\end{abstract}

\pacs{}


\maketitle 

\section{Introduction}\label{Sec:Intro}
Interacting regions of concentrated vorticity play a central role in the dynamics of a vast many fluid systems.
A classic example of such interactions is the vortex tripole \cite{vanheijst1991jfm}, in which three vortices move around each other for an extended period of time.  A collection of three vortices is arguably the most fundamental of the vortex configurations, as it contains the smallest number of vortices capable of exhibiting relative motion.
For modeling purposes, three point vortices in the plane, with positions $(x_{\alpha}, y_{\alpha})$ and circulations $\Gamma_{\alpha}$  (for $\alpha = 1,2,3$), provide the simplest reduced-order representation of a three-vortex system.  
A number of fundamental observations regarding vortex motion can be ascertained by considering the dynamics of three point vortices in the plane \cite{gro1877, syn1949, nov1975, are1979}.  

The motion of three interacting point vortices in the plane was solved by Gr\"{o}bli in his thesis of 1877 \cite{gro1877}.
More than 70 years later his solutions were classified in terms of the vortex circulations (or strengths) by Synge~\cite{syn1949} using a geometrical approach.
This work showed that the various possible regimes of motion are largely determined by the signs of the three symmetric functions of the circulations 
of the three interacting vortices, namely:
\begin{equation}
\begin{split}
&\gamma_1 = \Gamma_1+\Gamma_2+\Gamma_3,\\
&\gamma_2 = \Gamma_1\Gamma_2+\Gamma_2\Gamma_3+\Gamma_3\Gamma_1,\\
&\gamma_3 = \Gamma_1\Gamma_2\Gamma_3.
\end{split}\label{Eq:defgammas}
\end{equation}
While Gr\"{o}bli's approach leads to detailed solutions of the initial value problem, typically in terms of elliptic or hyperelliptic functions, Synge's approach is more in line with the qualitative, geometrical methods of modern dynamical systems theory.
The two approaches intersect for the case of identical vortices, where Gr\"{o}bli discusses a geometrical solution that is closely related to Synge's method for the general case.
Both the work of Gr\"{o}bli and of Synge lay dormant until Novikov revived interest in the three-vortex problem in 1975~\cite{nov1975}.
Novikov re-discovered, independently and yet almost word for word, the geometrical solution to the problem of three identical vortices that Gr\"{o}bli had given almost a century before.
Generalizing Novikov's work, Aref re-discovered Synge's geometrical approach for general vortex strengths, albeit in a slightly different form~\cite{are1979}.
The contorted history of re-discovery of solutions, and some background on the little known career of Gr\"{o}bli, has been recounted elsewhere~\cite{ART1992}.

In each of these previous geometrical solutions, the focus has been on describing the evolution of the vortex triangle in terms of the lengths of the sides, $s_1$, $s_2$, $s_3$, and the area,~$\Delta$.  
Alternatively, the geometry of the vortex triangle can be given in terms of the interior angles and the properties of the  circumcircle that passes through the vortex locations~\cite{are2002}.  In Ref.~\onlinecite{are2002}, this alternative geometrical description was used merely to arrive at a concise derivation of the existing equations for the lengths of the triangle sides. 
In Sec.~\ref{SS:RdotandZdot} and \ref{SS:Shapechange} we develop and explore the autonomous dynamical system given by the evolution of the circumcircle and the interior angles. 
In Sec.~\ref{SS:integrals} we discuss how the integrals of motion, including the Hamiltonian for the three vortex system, can be written purely in terms of the geometric variables. In Sec.~{\ref{SS:Z-R-relations}} we derive and discuss simple equations relating the center and radius of the circumcircle through the constants of motion, which are valid 
throughout the dynamical evolution of the system. In Sec.~\ref{sec:special-sol} we retrieve some of the well known results in three vortex motion through simple application of the equations of motion for the geometrical variables.

\section{Basic equations}
The motion of three vortices in the plane is given by six coupled, nonlinear, first order ODEs, two for the cartesian coordinates of each of the vortices, 
$(x_\alpha,y_\alpha)$, $\alpha = 1, 2, 3$.
If we concatenate the coordinates into complex positions $z_\alpha = x_\alpha + {\rm i}y_\alpha$, we have the equations of motion of these complex positions as:
\begin{subequations}\label{Eq:dzdt}
\begin{equation}
\overline{\frac{{\rm d}z_1}{{\rm d}t}} =
\frac{1}{2\pi{\rm i}}\left(\frac{\Gamma_2}{z_1-z_2}+ \frac{\Gamma_3}{z_1-z_3}\right),
\end{equation}
\begin{equation}
\overline{\frac{{\rm d}z_2}{{\rm d}t}} =
\frac{1}{2\pi{\rm i}}\left(\frac{\Gamma_1}{z_2-z_1}+ \frac{\Gamma_3}{z_2-z_3}\right),
\end{equation}
\begin{equation}
\overline{\frac{{\rm d}z_3}{{\rm d}t}} =
\frac{1}{2\pi{\rm i}}\left(\frac{\Gamma_1}{z_3-z_1}+ \frac{\Gamma_2}{z_3-z_2}\right),
\end{equation}
\end{subequations}
where the overbar denotes complex conjugation. We shall assume the basic equations (\ref{Eq:dzdt}) to be known.
For background 
on the information presented in this section we refer the reader to the textbook and monograph literature \cite{lam1932,bat1967,saf1992,new2001}.

Equations (\ref{Eq:dzdt}) have a number of well-known integrals.
Two of these are the components of linear impulse,
\begin{equation}
\label{Eq:QPdef}
Q = \Gamma_1x_1 + \Gamma_2x_2 + \Gamma_3x_3, \qquad
P = \Gamma_1y_1 + \Gamma_2y_2 + \Gamma_3y_3,
\end{equation}
which pertain to the absolute positions of the vortices.
The linear impulse determines the center of vorticity, 
\begin{equation}
z_{cv} = \frac{Q + {\rm i}P}{\gamma_1}.\label{Eq:zcvdef}
\end{equation}
For $\gamma_1 \ne 0$,
the origin may be shifted to $z_{cv}$ under the coordinate transformation $z_{\alpha} = z_{cv} + z'_{\alpha}$.  The linear impulse with respect to the center of vorticity is clearly zero, i.e. 
\begin{equation*}
	\Gamma_1 z'_1 + \Gamma_2 z'_2 + \Gamma_3 z'_3 = 0.
\end{equation*}

Another integral of (\ref{Eq:dzdt}) is the angular impulse,
\begin{subequations}\label{Eg:genIdef}
\begin{equation}\label{Eq:Idef}
I_0 = \Gamma_1(x^2_1 + y^2_1) + \Gamma_2(x^2_2 + y^2_2)  + \Gamma_3(x^2_3 + y^2_3),
\end{equation}
which also by definition requires the absolute positions of the vortices.
The subscript 0 signifies that $I_0$  is computed relative to the chosen origin of coordinates;
the angular impulse with respect to an arbitrary point $z$ can be written as 
\begin{equation}\label{Eq:Izdef}
	I_{z} = \sum_{\alpha=1}^{3} \Gamma_{\alpha} \left| z_{\alpha}- z \right|^2.
\end{equation}
\end{subequations}
One can also employ the above coordinate transformation to arrive at a \emph{parallel axis theorem} for the angular impulse, 
\begin{equation}\label{Eq:parallelthrm}
		I_{z} = I_{cv} + \gamma_1 \left| z - z_{cv} \right|^2,
\end{equation}
where $I_{cv}$ is the angular impulse calculated relative to the center of vorticity~(\ref{Eq:zcvdef}). 

The angular impulse is related to the quantity
\begin{equation}\label{Eq:defL}
L= \Gamma_1\Gamma_2s^2_3 + \Gamma_2 \Gamma_3s^2_1 + \Gamma_3 \Gamma_1s^2_2 =
\gamma_1I_0-Q^2-P^2.
\end{equation}
If we define the angular impulse with respect to $z_{cv}$, we  have $L=\gamma_1I_{cv}$.

Gr\"{o}bli discovered that one can isolate within this system of six real first-order ODEs a subsystem of three ODEs for the sides of the vortex triangle, $s_{\alpha}$, defined by
\begin{equation}
\begin{split}
&s^2_1 = |z_2-z_3|^2 = (x_2-x_3)^2 + (y_2-y_3)^2,\\
&s^2_2 = |z_3-z_1|^2= (x_3-x_1)^2 + (y_3-y_1)^2,\\
&s^2_3 = |z_1-z_2|^2= (x_1-x_2)^2 + (y_1-y_2)^2.
\end{split}\label{Eq:sidedefs}
\end{equation}
These equations are~\cite{gro1877}
\begin{subequations}
\begin{equation}
\frac{{\rm d}s^2_1}{{\rm d}t} =  \frac{2 \Delta}{\pi} \Gamma_1\frac{s^2_3 - s^2_2}{s^2_2s^2_3},\qquad
\frac{{\rm d}s^2_2}{{\rm d}t} =  \frac{2 \Delta}{\pi} \Gamma_2 \frac{s^2_1 - s^2_3}{s^2_3s^2_1},\qquad
\frac{{\rm d}s^2_3}{{\rm d}t} =  \frac{2 \Delta}{\pi} \Gamma_3 \frac{s^2_2 - s^2_1}{s^2_1s^2_2},
\label{Eq:Relmotion}
\end{equation}
where the area of the vortex triangle, $\Delta$, is given by {\it Heron's formula} 
\cite{cox1969},
\begin{equation}
16\Delta^2 = 2s^2_2s^2_3 + 2s^2_3s^2_1 + 2s^2_1s^2_2 - s^4_1 - s^4_2 - s^4_3.\label{Eq:Heron}
\end{equation}


The sign of $\Delta$ needed in (\ref{Eq:Relmotion}) is indeterminate from (\ref{Eq:Heron}) since a triangle and its mirror image have the same absolute area.
One needs to ``step outside'' the reduced system (\ref{Eq:Relmotion}) whenever the vortices become collinear and consider the triangle area with orientation,
\begin{equation}
\Delta = \textstyle \frac{1}{2}(x_1y_2 +x_2y_3 + x_3y_1 - x_1y_3 -x_3y_2 -x_2y_1).\tag{\ref{Eq:Heron}$'$}
\end{equation}
This definition makes $\Delta>0$ when vortices 123 appear counter-clockwise and $\Delta<0$ when they appear clock-wise.
The evolution of $\Delta$ may now be traced via (\ref{Eq:dzdt}).
The subsystem (\ref{Eq:Relmotion},\ref{Eq:Heron}), then, is ``closed'' except for instants when the vortices become collinear.
In order to know how to continue the motion through such instants, one needs to either appeal to the full equations of motion 
(\ref{Eq:dzdt}) or develop a fourth equation of motion for $\Delta$ itself in terms of $s_1, s_2, s_3$.
This equation is~\cite{BP1998}
\begin{equation}
\frac{{\rm d}\Delta}{{\rm d}t} =
\frac{1}{8\pi}\bigg[(\Gamma_1 +
\Gamma_2)\frac{s^2_1 - s^2_2}{s^2_3}\ + 
(\Gamma_2 + \Gamma_3)\frac{s^2_2 -s^2_3}{s^2_1} + (\Gamma_3 + \Gamma_1 )\frac{s^2_3 -s^2_1}{s^2_2}\bigg].
\label{Eq:dDeltadt}
\end{equation}
\end{subequations}

We do not include derivations of either (\ref{Eq:Relmotion}) or (\ref{Eq:dDeltadt}) in this paper, but the solution procedure is as follows.
Given equations for the velocities of the three corners of a triangle, it is clear, in principle, that one can write equations for the 
time-rate-of-change of the triangle sides and the triangle area.
The geometrical considerations required for such a ``direct'' derivation are, however, somewhat involved.
A straightforward algebraic development can be achieved by appealing to the Hamiltonian formulation~\cite{are1983} of (\ref{Eq:dzdt}) 
introduced already by Kirchhoff in 1876 and well covered in the texts cited \cite{lam1932, bat1967,saf1992,new2001}.
If we set
\begin{equation}
H = - \frac{1}{4\pi} (\Gamma_1\Gamma_2\log s^2_3 + \Gamma_2\Gamma_3 \log s^2_1 + \Gamma_3\Gamma_1\log s^2_2),\label{Eq:ThreevortexH}
\end{equation}
it is not difficult to verify that (\ref{Eq:dzdt}) may be written
\begin{equation}
\Gamma_\alpha\frac{{\rm d}x_\alpha}{{\rm d}t} = \frac{\partial H}{\partial y_\alpha},\quad
\Gamma_\alpha\frac{{\rm d}y_\alpha}{{\rm d}t} = -\frac{\partial H}{\partial x_\alpha} \quad (\alpha = 1,2,3).
\label{Eq:Hamileqs}
\end{equation}
Thus, $x_\alpha$ and $\Gamma_\alpha y_\alpha$ are canonically conjugate variables, and one can introduce a Poisson bracket~\cite{LL1960,GPS2002}
\begin{equation}
[f,g] = \sum^3_{\alpha =1}\frac{1}{\Gamma_\alpha} \left(\frac{\partial f}{\partial x_\alpha} \frac{\partial g}{\partial y_\alpha} - \frac{\partial f}{\partial y_\alpha} \frac{\partial g}{\partial x_\alpha}\right).\label{Eq:PB}
\end{equation}
In the context of point vortex dynamics this development dates back at least to a 1905 paper by Laura~\cite{lau1905}.
From the fundamental Poisson brackets,
\begin{equation}
[x_1,\Gamma_1y_1] = [x_2,\Gamma_2y_2] =  [x_3,\Gamma_3y_3]  = 1,\label{Eq:[x,Gy]}
\end{equation}
with all other brackets of two coordinates equal to zero,
one builds up the algebra to produce results such as~\cite{BP1998}
\begin{equation}
[s^2_1,s^2_2] = -8\frac{\Delta}{\Gamma_3},\qquad
[s^2_2,s^2_3] = -8\frac{\Delta}{\Gamma_1},\qquad
[s^2_3,s^2_1] = -8\frac{\Delta}{\Gamma_2}.
\label{Eq:PBs2}
\end{equation}
Then, since the evolution of any function of the coordinates is given by
\begin{equation}
\frac{{\rm d}f}{{\rm d}t} = \frac{\partial f}{\partial t} + [f,H],\label{Eq:dfdt}
\end{equation}
and $H$ is given by (\ref{Eq:ThreevortexH}), one finds (\ref{Eq:Relmotion}) and (\ref{Eq:dDeltadt}) after straightforward  calculation of Poisson brackets.

\section{Evolution of vortex triangle geometry}

One can show~\cite{are2002} that the vortex velocities may be written quite simply in terms of the interior 
angles of the vortex triangle and the radius of the circumscribed circle, $R$.
These results allow us to derive equations of motion for $R$ and for the velocity of the center of the circumcircle, $Z$.
If the angles of the triangle are denoted $A, B, C$ as shown in Fig.~\ref{fig:defAngles}, 
then we have the geometrical relations 
\begin{equation}
s_1 = 2R \sin A,\qquad
s_2 = 2R \sin B,\qquad
s_3 = 2R \sin C.\label{Eq:s=Rsinangle}
\end{equation}
For later reference, we also note the relation
\begin{equation}
R = \frac{s_1s_2s_3}{4|\Delta|}\label{Eq:Rfromsides}
\end{equation}
and the fundamental relation between the interior angles
\begin{equation}\label{Eq:interiorangles}
	A + B + C = \pi.
\end{equation}
These relations allow us to derive equations for the evolution of the triangle shape.

When the vortices become collinear, both the radius and the center of the circumcircle go to infinity,  
and the positions of the vortices in terms of the circumcircle 
become ill-defined.
Since this geometrical approach breaks down as the vortices pass through a collinear state, 
we can choose, without any
loss of generality, to always label the vortices  such that they are oriented counterclockwise, which gives $\Delta>0$ in all of the subsequent analysis.

\subsection{Evolution of the radius and center of the circumcircle }\label{SS:RdotandZdot}

\begin{figure}
	\includegraphics[width=\textwidth]{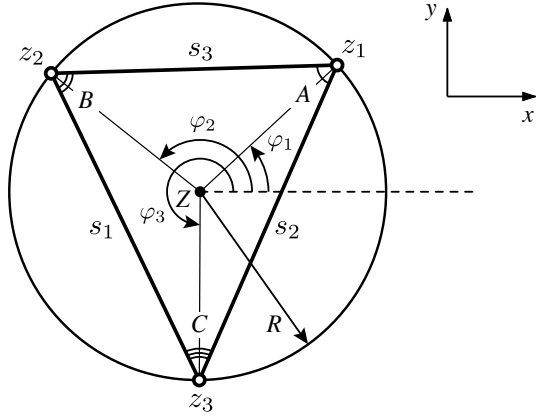}
	\caption{Definition of the geometrical variables.  Open circles mark the vortex locations $z_1, z_2, z_3$. Filled circle marks the center of the circumcircle, $Z$.}
	\label{fig:defAngles}
\end{figure}

As shown in Fig.~\ref{fig:defAngles}, the vortex co-ordinates can be written as 
\begin{equation}\label{Eq:zfromZRphi}
z_1 = Z + R{\rm e}^{{\rm i}\varphi_1},\qquad
z_2 = Z + R{\rm e}^{{\rm i}\varphi_2},\qquad
z_3 = Z + R{\rm e}^{{\rm i}\varphi_3},
\end{equation}
where $\varphi_\alpha$ measures the angle made by the position vector of vortex $\alpha$ with respect to the $x$ (horizontal) axis.
If the three vortices appear counterclockwise, as we have assumed, it follows from elementary geometry that the interior angles of the vortex triangle  are given by
\begin{equation*}
\varphi_2 - \varphi_1 = 2C,\qquad
\varphi_3 - \varphi_2 = 2A,\qquad
\varphi_1 - \varphi_3 = 2B - 2\pi.
\end{equation*}

With this notation, we can write the equation of motion for, e.g., vortex 1 as
\begin{equation*}
\dot{\overline z}_1 =
\frac{1}{2\pi {\rm i} R} \bigg(\frac{\Gamma_2}{{\rm e}^{{\rm i}\varphi_1} - {\rm e}^{{\rm i}\varphi_2}} + \frac{\Gamma_3}{{\rm e}^{{\rm i}\varphi_1} - {\rm e}^{{\rm i}\varphi_3}}  \bigg)=
\frac{{\rm e}^{-{\rm i}\varphi_1}}{2\pi {\rm i} R}\bigg(\frac{\Gamma_2}{1 - {\rm e}^{{\rm i}2C}} + \frac{\Gamma_3}{1 - {\rm e}^{-{\rm i}2B}}  \bigg),
\end{equation*}
where the overdot denotes the time-derivative.  This expression may be written as
\begin{equation}\label{Eq:z1dot}
\dot{z}_1 =
\frac{{\rm e}^{{\rm i}\varphi_1}}{4\pi R}\big[\Gamma_2\cot C - \Gamma_3\cot B + {\rm i}(\Gamma_2 + \Gamma_3) \big].
\end{equation}
The real multiple of ${\rm e}^{{\rm i}\varphi_1}$ in \eqref{Eq:z1dot} gives the radial velocity of vortex 1 relative to the circumcircle, namely 
\begin{subequations}\label{Eq:z1radtan}
\begin{equation}\label{Eq:z1rad}
(\dot{z}_1)_{rad} = \frac{\Gamma_2\cot C - \Gamma_3\cot B}{4\pi R}.
\end{equation}
The real multiple of ${\rm i}{\rm e}^{{\rm i}\varphi_1}$  in \eqref{Eq:z1dot} gives the tangential velocity in the positive (counterclockwise) direction, namely 
\begin{equation}\label{Eq:z1tan}
(\dot{z}_1)_{tan} = \frac{\Gamma_2 + \Gamma_3}{4\pi R}.
\end{equation}
\end{subequations}


Let the center of the circumcircle be $Z = X + {\rm i}Y$. We may then equate $(\dot{z}_1)_{rad}$ in equation~\eqref{Eq:z1rad} to the sum of 
the rate of change of $R$ 
plus
the projection of $\dot Z$ onto the radial direction from $Z$ to $z_1$, giving
\begin{subequations}\label{Eq:z1vel}
\begin{equation}
\dot R + \dot X \cos\varphi_1 + \dot Y\sin\varphi_1 =  \frac{\Gamma_2\cot C - \Gamma_3\cot B}{4\pi R}.
\label{Eq:radvel}
\end{equation}
Similarly, we may equate $(\dot{z}_1)_{tan}$  in equation~\eqref{Eq:z1tan} to the sum of $R\,\dot\varphi_1$ and the tangential component of $\dot Z$ at the position of vortex 1, giving
\begin{equation}
R\,\dot\varphi_1  -\dot X \sin\varphi_1 + \dot Y\cos\varphi_1  = \frac{\Gamma_2 + \Gamma_3}{4\pi R}.
\label{Eq:tanvel}
\end{equation}
\end{subequations}
Alternatively, we may simply differentiate (\ref{Eq:zfromZRphi}) to find that
\begin{equation*}
{\rm e}^{-{\rm i}\varphi_1}\dot{z}_1 = {\rm e}^{-{\rm i}\varphi_1}\dot Z + \dot R + {\rm i}R\dot\varphi_1.
\end{equation*}
The real and imaginary parts of this equation give (\ref{Eq:radvel}) and (\ref{Eq:tanvel}), respectively.
Similar relations hold for the velocity components of vortices 2 and 3.

From equations~\eqref{Eq:z1vel} and the corresponding equations for vortices 2 and 3, 
we have a system of equations that may  be written in matrix form 
%
using the Cartesian coordinates of the vortices as
\begin{equation}
\begin{bmatrix}
\dot{X} \\
\dot{Y} \\
\dot{R}
\end{bmatrix} = \ \frac{1}{8\pi \Delta} 
\begin{bmatrix}
		y_2-y_3 & y_3-y_1 & y_1-y_2 \\
		x_3-x_2 & x_1-x_3 & x_2-x_1 \\
		R\sin(2A) & R\sin(2B) & R\sin(2C) 
\end{bmatrix}
\begin{bmatrix}
		\Gamma_2\cot C - \Gamma_3\cot B \\
		\Gamma_3\cot A - \Gamma_1\cot C\\
		\Gamma_1\cot B - \Gamma_2\cot A
\end{bmatrix},
\label{Eq:Matrixeqs}
\end{equation}
where we have used the assumption that $\Delta > 0$.
%
To write the equation for $\dot R$ in its most transparent form, we collect terms proportional 
to each of the circulations $\Gamma_1$, $\Gamma_2$ and $\Gamma_3$, giving
\begin{subequations}\label{Eq:Rdot}
\begin{equation}
\begin{split}
\dot R = & \frac{1}{16\pi R\sin A \sin B  \sin C}\  \Biggl\{ \Gamma_1[\cot B\sin(2C) - \cot C\sin(2B)]\ \\
&\quad + \Gamma_2[\cot C\sin(2A) - \cot A\sin(2C)]\ + \Gamma_3[\cot A\sin(2B) - \cot B\sin(2A)] \Biggr\},
\end{split}
\end{equation}
or, again using equation~\eqref{Eq:interiorangles},
\begin{equation}
\begin{split}
	\frac{{\rm d}R^2}{{\rm d}t} = \frac{1}{4\pi} \Bigl[ &\Gamma_1\cot B\cot C(\cot B - \cot C)\ \\
+ & \ \Gamma_2\cot C\cot A(\cot C - \cot A) + \Gamma_3\cot A\cot B(\cot A - \cot B)\Bigr].
\end{split}\label{Eq:Rdotfinal}
\end{equation}
\end{subequations}
For the motion of the center of the circumcircle we find from (\ref{Eq:Matrixeqs}) that
\begin{subequations}
\begin{equation}\label{Eq:EoMZ}
\begin{split}
\dot{Z} = \dot X + {\rm i}\dot Y=\ & \frac{1}{8\pi {\rm i}\Delta}\ \Bigl[(z_1-z_2)(\Gamma_1\cot B - \Gamma_2\cot A)  \\
	&\ + (z_2-z_3)(\Gamma_2\cot C - \Gamma_3\cot B) +  (z_3-z_1)(\Gamma_3\cot A - \Gamma_1\cot C)\Bigr].
\end{split}
\end{equation}
By re-grouping terms, 
we find
\begin{equation}\label{Eq:EoMZ1}
\dot{Z} = \frac{1}{8\pi {\rm i}\Delta}\Bigl[(Q + {\rm i}P)(\cot A + \cot B + \cot C)- \gamma_1(z_1\cot A  + z_2\cot B  + z_3\cot C)\Bigr],
\end{equation}
where $\gamma_1$ is given in \eqref{Eq:defgammas}.
\end{subequations}
The advantage of using this form of the equation is that, if $\gamma_1 = 0$,  the second term 
in square brackets vanishes; if instead $\gamma_1 \ne 0$, we can arrange for the center of vorticity (\ref{Eq:zcvdef})
to be at the origin, in which case $Q+{\rm i}P=0$.
Note that since $A+B+C=\pi$~\eqref{Eq:interiorangles},
\begin{equation*}
\cot A + \cot B + \cot C = {\textstyle\frac{1}{2}}(\sin^2A + \sin^2B + \sin^2C)\ge0,
\end{equation*}
with the equality occurring only when the vortices are collinear.

\subsection{Evolution of triangle shape}\label{SS:Shapechange}

Next, we seek equations of motion for the interior angles $A, B, C$.
From (\ref{Eq:s=Rsinangle}), we have
\begin{equation}
\cot A \frac{{\rm d}A}{{\rm d}t} = \frac{1}{2R\sin A} \frac{{\rm d}s_1}{{\rm d}t} - \frac{1}{R}\frac{{\rm d}R}{{\rm d}t}.\label{Eq:Intermediate}
\end{equation}
Here, from (\ref{Eq:Relmotion}),
\begin{equation}
\begin{split}
\frac{{\rm d}s_1}{{\rm d}t} =&\ \frac{\Gamma_1\Delta}{\pi s_1}\left[\frac{s^2_3 - s^2_2}{s^2_3s^2_2} \right]
= \frac{\Gamma_1}{8\pi R}\left[\frac{\cos(2B) - \cos(2C)}{\sin B\sin C}\right]\\
=& \frac{\Gamma_1}{4\pi R}\sin A (\cot B - \cot C).
\end{split}\label{Eq:ds1dt}
\end{equation}
%
%
Combining (\ref{Eq:Rdotfinal}), (\ref{Eq:Intermediate}) and (\ref{Eq:ds1dt}) gives
\begin{subequations}\label{Eq:dAdt}
\begin{equation}
\begin{split}
\cot A \frac{{\rm d}A}{{\rm d}t} =&\frac{1}{8\pi R^2}\ 
\Big[\Gamma_1 (1-\cot B \cot C) (\cot B-\cot C)\ -\\
\frac{}{}&\Gamma_2 \cot C \cot A (\cot C-\cot A)\ -
\Gamma_3 \cot A \cot B (\cot A-\cot B)\Big].
\end{split}
\end{equation}
For completeness we write out the corresponding equations for ${\rm d}B/{\rm d}t$ and ${\rm d}C/{\rm d}t$:
\begin{equation}
\begin{split}
\cot B \frac{{\rm d}B}{{\rm d}t} =&\frac{1}{8\pi R^2}\ 
\Big[\Gamma_2 (1-\cot C \cot A) (\cot C-\cot A)\ -\\
\frac{}{}&\Gamma_3 \cot A \cot B (\cot A-\cot B)\ -
\Gamma_1 \cot B \cot C (\cot B-\cot C)\Big],
\end{split}\label{Eq:dBdt}
\end{equation}
\begin{equation}
\begin{split}
\cot C \frac{{\rm d}C}{{\rm d}t} =&\frac{1}{8\pi R^2}\
\Big[\Gamma_3 (1-\cot A \cot B) (\cot A-\cot B)\ -\\
\frac{}{}&\Gamma_1 \cot B \cot C (\cot B-\cot C)\ -
\Gamma_2 \cot C \cot A (\cot C-\cot A)\Big].
\end{split}\label{Eq:dCdt}
\end{equation}
\end{subequations}
Equations (\ref{Eq:Rdotfinal}) and (\ref{Eq:dAdt}) form an autonomous four-dimensional dynamical system embedded in the six-dimensional system (\ref{Eq:dzdt}).

There is, in addition, the obvious constraint on
the three angles 
\eqref{Eq:interiorangles},
so the system consisting of (\ref{Eq:Rdotfinal}) and (\ref{Eq:dAdt}) may 
be thought of as three-dimensional. We saw earlier that the system comprised of (\ref{Eq:Relmotion}) and 
(\ref{Eq:dDeltadt}) was also three-dimensional because of (\ref{Eq:Heron}), except for those instants when the three vortices become collinear.
The role of collinear configurations shows up in a different way in the system consisting of (\ref{Eq:Rdotfinal}) with (\ref{Eq:dAdt}):
Collinear configurations are singularities of these equations wherein $R\rightarrow\infty$.
It is thus clear that one has to stop and consider how to continue the solution beyond such a singularity in this formulation.

\subsection{Integrals of motion}\label{SS:integrals}
We return to a consideration of the integrals of motion.
In the reduced description provided by (\ref{Eq:Relmotion}), conservation of $L$ (\ref{Eq:defL}) follows immediately by dividing 
the left hand sides in (\ref{Eq:Relmotion}) by $\Gamma_1$, $\Gamma_2$ and $\Gamma_3$, respectively, and adding.
The Hamiltonian (\ref{Eq:ThreevortexH}) is also an integral of the motion.
This again follows easily from (\ref{Eq:Relmotion}) by dividing the left hand sides by $\Gamma_1s^2_1$, $\Gamma_2s^2_2$ and $\Gamma_3s^2_3$, respectively, and adding.

In terms of the variables $R, A, B, C$ we have from (\ref{Eq:s=Rsinangle}) that
\begin{subequations}\label{Eq:com}
\begin{equation}
L = 4\gamma_3R^2\left(\frac{\sin^2A}{\Gamma_1} +\frac{\sin^2B}{\Gamma_2} +\frac{\sin^2C}{\Gamma_3}\right)\label{Eq:LRABC}
\end{equation}
and
\begin{equation}
H = -\frac{1}{2\pi}\Bigg[\gamma_2\log R\ + \gamma_3\left(\frac{\log\sin A}{\Gamma_1} + \frac{\log\sin B}{\Gamma_2}  + \frac{\log\sin C}{\Gamma_3}\right) \Bigg].
\label{Eq:HamilRABC}
\end{equation}
\end{subequations}
These must be integrals of (\ref{Eq:Rdotfinal}) and (\ref{Eq:dAdt}), as may be verified directly.
The verification, however, takes a few steps of not entirely transparent algebra, and one may wonder if these integrals would have been discovered working within the dynamical system (\ref{Eq:Rdotfinal}), (\ref{Eq:dAdt}) without the general background we have given.
We leave the details to the reader.

In addition we have a purely geometrical integral, {\it viz} $A + B + C = \pi$ \eqref{Eq:interiorangles}.
That the sum $A + B + C$ has vanishing time derivative also follows by adding the equations of motion for $A$, $B$ and $C$, (\ref{Eq:dAdt}), and noting that the net coefficient of each of $\Gamma_1$, $\Gamma_2$ and $\Gamma_3$ vanishes.
Here one needs to use relations such as
\begin{equation*}
\frac{1}{\cot A} = \frac{\cot B + \cot C}{1-\cot B\cot C},
\end{equation*}
which holds when $A + B + C = \pi$.

\subsection{EoM for the center of the circumcircle revisited}
We have already derived the equation of motion for the center of the circumcircle $Z$.
In this subsection we show how that result can be obtained using the canonical formalism.
In the next subsection we then show how the equation of motion for $R$ follows from the equation of motion for $Z$.

We return to the coordinates of the vortices and note that (\ref{Eq:Heron}) is equivalent to
\begin{equation}
4{\rm i}\Delta = \overline{z}_1(z_2-z_3) + \overline{z}_2(z_3-z_1) + \overline{z}_3(z_1-z_2).\label{Eq:Deltacomplex}
\end{equation}
From the fundamental Poisson brackets (\ref{Eq:[x,Gy]}), which in terms of the complex vortex coordinates read
\begin{equation}
[z_\alpha,z_\beta] = 0,\qquad
[z_\alpha, \overline{z}_\beta] = -\frac{2{\rm i}}{\Gamma_\alpha}\delta_{\alpha\beta},
\end{equation}
it follows that
\begin{subequations}
\begin{align}
&[z_1,\Delta] = -\frac{z_2-z_3}{2\Gamma_1},\label{Eq:PBzDeltaa}\\
&[z_2,\Delta] = -\frac{z_3-z_1}{2\Gamma_2},\label{Eq:PBzDeltab}\\
&[z_3,\Delta] = -\frac{z_1-z_2}{2\Gamma_3}.\label{Eq:PBzDeltac}
\end{align}
\end{subequations}

Next, we pause to derive a well known 
expression for the center of the circumcircle in terms of the positions of the vertices of the triangle.
By definition of the circumcircle and its radius we have $|z_1 - Z| = |z_2 - Z| = |z_3 - Z| = R$.
Thus,
\begin{equation}
\begin{split}
&z_1\overline{z}_1 + |Z|^2 - z_1\overline{Z} - \overline{z}_1Z = R^2,\\
&z_2\overline{z}_2 + |Z|^2 - z_2\overline{Z} - \overline{z}_2Z = R^2,\\
&z_3\overline{z}_3 + |Z|^2 - z_3\overline{Z} - \overline{z}_3Z = R^2.
\end{split}\label{Eq:|z-Z|=R}
\end{equation}
By eliminating $|Z|^2-R^2$ from these relations we obtain
\begin{subequations}
\begin{align}
&(\overline{z}_1 - \overline{z}_2)Z + (z_1-z_2)\overline{Z} = z_1\overline{z}_1 - z_2\overline{z}_2,\label{Eq:derivZa}\\
&(\overline{z}_2 - \overline{z}_3)Z + (z_2-z_3)\overline{Z} = z_2\overline{z}_2 - z_3\overline{z}_3,
\end{align}
\end{subequations}
and solving these two linear equations for $Z$ and $\overline{Z}$ we find
\begin{equation}
Z = 
\frac{|z_1|^2(z_2-z_3) + |z_2|^2(z_3-z_1) + |z_3|^2(z_1-z_2)}{\overline{z}_1(z_2-z_3) + \overline{z}_2(z_3-z_1) + \overline{z}_3(z_1-z_2)}.
\label{Eq:Zfromzs}
\end{equation}
From (\ref{Eq:Deltacomplex}) the denominator is seen to be $4{\rm i}\Delta$.

Since we have equations of motion for the vortex positions and for $\Delta$ (\ref{Eq:dDeltadt}), we can, in principle, derive an equation of motion for $Z$ from (\ref{Eq:Zfromzs}).
In order to do so we calculate the Poisson bracket $[Z,H]$.
Using (\ref{Eq:PBzDeltaa}) we get
\begin{equation*}
[z_1,4{\rm i}\Delta Z] = 4{\rm i}([z_1,\Delta] Z + [z_1,Z]\Delta) =
4{\rm i}\left(\frac{z_3-z_2}{2\Gamma_1}Z + [z_1,Z]\Delta\right).
\end{equation*}
However, from the fundamental Poisson brackets, and in view of (\ref{Eq:Zfromzs}), the left hand side is clearly
\begin{equation*}
[z_1,4{\rm i}\Delta Z] = [z_1,|z_1|^2(z_2-z_3)] = z_1(z_3-z_2)\frac{2{\rm i}}{\Gamma_1}.
\end{equation*}
Thus,
\begin{subequations}
\begin{equation}
[z_1,Z] =
-\frac{(z_2-z_3)(z_1-Z)}{2\Delta\Gamma_1},\label{Eq:PBz1Z}
\end{equation}
and by permutation of indices,
\begin{equation}
[z_2,Z] = -\frac{(z_3-z_1)(z_2-Z)}{2\Delta\Gamma_2},\label{Eq:PBz2Z}
\end{equation}
\begin{equation}
[z_3,Z] = -\frac{(z_1-z_2)(z_3-Z)}{2\Delta\Gamma_3}.\label{Eq:PBz3Z}
\end{equation}
\end{subequations}
As a corollary
\begin{equation}
[\Gamma_1z_1+\Gamma_2z_2 + \Gamma_3z_3,Z] = 0.
\end{equation}

Taking the Poisson bracket of (\ref{Eq:derivZa}) with $\overline{z}_3$, we get
\begin{equation*}
(\overline{z}_1 - \overline{z}_2)[\overline{z}_3,Z] + (z_1-z_2)[\overline{z}_3,\overline{Z}]=0.
\end{equation*}
Here $[\overline{z}_3,\overline{Z}] = \overline{[z_3,Z]}$, since the Poisson bracket operations are all in terms of real-valued quantities.
Thus,
\begin{equation*}
(\overline{z}_1 - \overline{z}_2)[\overline{z}_3,Z] - (z_1-z_2)\frac{(\overline{z}_1-\overline{z}_2)(\overline{z}_3-\overline{Z})}{2\Delta\Gamma_3} =0,
\end{equation*}
or
\begin{subequations}
\begin{equation}
[\overline{z}_3,Z] = \frac{(z_1-z_2)(\overline{z}_3-\overline{Z})}{2\Delta\Gamma_3}.\label{Eq:PBz3barZ}
\end{equation}
By permutation of indices,
\begin{equation}
[\overline{z}_2,Z] = \frac{(z_3-z_1)(\overline{z}_2-\overline{Z})}{2\Delta\Gamma_2},\label{Eq:PBz2barZ}
\end{equation}
\begin{equation}
[\overline{z}_1,Z] = \frac{(z_2-z_3)(\overline{z}_1-\overline{Z})}{2\Delta\Gamma_1}.\label{Eq:PBz1barZ}
\end{equation}
\end{subequations}
We are now in a position to calculate
\begin{equation*}
\begin{split}
\Gamma_1&\Gamma_2[Z,\log s^2_3] =
\frac{\Gamma_1\Gamma_2}{s^2_3}[Z,(z_1-z_2)(\overline{z}_1 - \overline{z}_2)] =\\
&\frac{\Gamma_1\Gamma_2}{s^2_3}\big([Z,z_1-z_2](\overline{z}_1 - \overline{z}_2) + [Z,\overline{z}_1 - \overline{z}_2](z_1-z_2)\big).
\end{split}
\end{equation*}
Here, by (\ref{Eq:PBz1Z}) and (\ref{Eq:PBz1barZ}),
\begin{equation*}
\begin{split}
[Z,&z_1](\overline{z}_1 - \overline{z}_2) + [Z,\overline{z}_1](z_1-z_2) =\\
&\frac{(z_2-z_3)}{2\Delta\Gamma_1}[(z_1-Z)(\overline{z}_1 - \overline{z}_2) -(\overline{z}_1-\overline{Z})(z_1-z_2)] =\\&-\frac{(z_2-z_3)(\overline{z}_1 - \overline{z}_2)}{\Delta\Gamma_1}[Z  -{\textstyle\frac{1}{2}}(z_1+z_2)],
\end{split}
\end{equation*}
where in the last step (\ref{Eq:derivZa}) has been used in the form
\begin{equation*}
(z_1-z_2)\overline{Z} = -(\overline{z}_1 - \overline{z}_2)Z + |z_1|^2 - |z_2|^2.
\end{equation*}
Thus,
\begin{equation*}
\begin{split}
[Z,&z_1](\overline{z}_1 - \overline{z}_2) + [Z,\overline{z}_1](z_1-z_2) =\\
&\frac{(z_2-z_3)}{2\Delta\Gamma_1}[2Z(\overline{z}_1 - \overline{z}_2)  - z_1\overline{z}_2 + \overline{z}_1z_2 + |z_1|^2 - |z_2|^2] =\\
&-\frac{(z_2-z_3)(\overline{z}_1 - \overline{z}_2)}{\Delta\Gamma_1}[Z  -{\textstyle\frac{1}{2}}(z_1+z_2)].
\end{split}
\end{equation*}
Similarly,
\begin{equation*}
\begin{split}
[Z,&z_2](\overline{z}_1 - \overline{z}_2) + [Z,\overline{z}_2](z_1-z_2) =\\
&\frac{(z_3-z_1)}{2\Delta\Gamma_2}[(z_2-Z)(\overline{z}_1 - \overline{z}_2) -(\overline{z}_2-\overline{Z})(z_1-z_2)] =\\&-\frac{(z_3-z_1)(\overline{z}_1 - \overline{z}_2)}{\Delta\Gamma_2}[Z  -{\textstyle\frac{1}{2}}(z_1+z_2)].
\end{split}
\end{equation*}
Subtracting we have
\begin{equation*}
\begin{split}
\Gamma_1&\Gamma_2[Z,\log s^2_3] =
\frac{\Gamma_1\Gamma_2(\overline{z}_1 - \overline{z}_2)}{\Delta s^2_3}\ \times\\
&\big[-\frac{(z_2-z_3)}{\Gamma_1}+ \frac{(z_3-z_1)}{\Gamma_2}\big][Z  -{\textstyle\frac{1}{2}}(z_1+z_2)] =\\
&\frac{\overline{z}_1 - \overline{z}_2}{\Delta s^2_3}[\gamma_1z_3 - (Q+{\rm i}P)][Z  -{\textstyle\frac{1}{2}}(z_1+z_2)] =\\
&\quad\frac{\gamma_1z_3 - (Q+{\rm i}P)}{\Delta}\ \frac{Z  -{\textstyle\frac{1}{2}}(z_1+z_2)}{z_1-z_2},
\end{split}
\end{equation*}
with similar terms from the other two terms in the Hamiltonian.

Now, in terms of the angles in the vortex triangle
\begin{equation*}
\begin{split}
&\frac{Z  -{\textstyle\frac{1}{2}}(z_1+z_2)}{z_1-z_2} =
-{\textstyle\frac{1}{2}}\frac{{\rm e}^{{\rm i}\varphi_1} + {\rm e}^{{\rm i}\varphi_2}}{{\rm e}^{{\rm i}\varphi_1} - {\rm e}^{{\rm i}\varphi_2}} =\\
&\quad -{\textstyle\frac{1}{2}}\frac{1+ {\rm e}^{2{\rm i}C}}{1 - {\rm e}^{2{\rm i}C}} =
\frac{1}{2{\rm i}}\frac{\cos C}{\sin C} =
{-\textstyle\frac{1}{2}}{\rm i}\cot C.
\end{split}
\end{equation*}
Thus,
\begin{equation*}
\begin{split}
-\frac{\Gamma_1\Gamma_2}{4\pi}[Z,\log s^2_3] =
\frac{(Q+{\rm i}P) - \gamma_1z_3}{8\pi{\rm i}\Delta}\cot C.
\end{split}
\end{equation*}
The contributions to $[Z,H]$ from the other two terms in the Hamiltonian follows by permutation of indices, and we obtain again (\ref{Eq:EoMZ1}):
\begin{equation*}
\begin{split}
\dot Z = [Z,H]=
&\frac{1}{8\pi{\rm i}\Delta}[(Q+{\rm i}P)(\cot A+\cot B + \cot C)-\\
&\qquad\qquad \gamma_1(z_1\cot A+z_2\cot B + z_3\cot C)\ ].
\end{split}
\end{equation*}
However, it is clear that the geometrical derivation is much simpler than the direct algebraic approach!

\subsection{EoM for the radius of the circumcircle revisited}\label{SS:Z-R-relations}
In Section~\ref{SS:RdotandZdot} we derived the equation of motion for $R$.
A different derivation is obtained by starting from the geometrical result (\ref{Eq:Rfromsides})
and using the equations of motion for the sides, (\ref{Eq:Relmotion}), and for the area, (\ref{Eq:dDeltadt}).  These may then be combined to produce (\ref{Eq:Rdotfinal}).

Here we pursue a somewhat different avenue.
We note the following relation between $Z$ and $R$:
Multiply the first of (\ref{Eq:|z-Z|=R}) by $\Gamma_1$, the second by $\Gamma_2$, and the third by $\Gamma_3$, and add the results.
This gives
\begin{subequations}
\begin{equation}\label{Eq:PAZzc}
I_0 + \gamma_1Z\overline{Z} - (Q+{\rm i}P)\overline{Z} - (Q-{\rm i}P)Z = \gamma_1R^2,
\end{equation}
where $Q$ and $P$ are as in (\ref{Eq:QPdef}), $I_0$ as in (\ref{Eq:Idef}).

If the sum of the vortex circulations, $\gamma_1$, vanishes, the projection of the vector from the origin to the circumcenter, $Z$, onto the (constant) linear impulse is constant.
Thus, the circumcenter travels along a line perpendicular to $Q+{\rm i}P$.
Analysis~\cite{rot1989,are1989} shows that the vortices periodically become collinear for all initial conditions.
As the vortices become collinear, $Z$ recedes to infinity 
along a line perpendicular to $Q+{\rm i}P$.
In other words, the collinear vortices are situated along $Q+{\rm i}P$.
If $Q=P=0$, the vortices must remain collinear and will, as the analysis shows~\cite{rot1989,are1989}, rotate like a rigid body.

The same conclusions are reached from (\ref{Eq:EoMZ1}).
For $\gamma_1=0$ in that equation, the second term in square brackets is absent and $\dot Z$ is an imaginary number times $Q+{\rm i}P$.

In the general case, $\gamma_1 \ne 0$, we may view (\ref{Eq:PAZzc}) as an example of the parallel axis theorem~(\ref{Eq:parallelthrm}).
Using (\ref{Eq:Izdef}) to write $I_Z = I_{cv} + \gamma_1|Z - z_{cv}|^2$, and combining this equation with (\ref{Eq:Idef}) allows us to
write (\ref{Eq:PAZzc}) as (for $\gamma_1 \ne 0$)
\begin{equation}\label{Eq:ZandR}
|Z - z_{cv}|^2 = R^2 - \frac{I_{cv}}{\gamma_1}.
\end{equation}
\end{subequations}
Since $z_{cv}$ and $I_{cv}$ are dynamical invariants, this equation shows that if we have determined the evolution of $Z$, we also have the evolution of $R$.
If we place $z_{cv}$ at the origin, which requires $\gamma_1 \ne 0$, we have that $|Z|^2 - R^2$ is a constant of the motion.

Equation (\ref{Eq:ZandR}) has an interesting geometrical interpretation.
By (\ref{Eq:defL}), $L=\gamma_1 I_{cv}$, and the sign of $I_{cv}/\gamma_1$ is the same as the sign of $L$. 
If $L=0$, (\ref{Eq:ZandR}) reduces to 
\begin{equation}
	|Z - z_{cv}|^2 = R^2.
\end{equation}
 Thus, while the circumcircle is, in general, time dependent, it evolves in such a way that one point on the circumcircle is always at the center of vorticity.

If $L>0$, then (\ref{Eq:ZandR}) can be written as
\begin{equation}
	|Z - z_{cv}|^2 + \left|\frac{I_{cv}}{\gamma_1}\right|= R^2,
	\label{Eq:ZandRLpos}
\end{equation}
and we see that the center of vorticity must lie inside the circumcircle for all times.
Furthermore, the circumradius $R$ has a lower bound, namely $R_{min}=\sqrt{{I_{cv}}/{\gamma_1}}$, which  
corresponds to the circumcenter coinciding with the center of vorticity. On the other hand, if $L<0$,
(\ref{Eq:ZandR}) may be written as
\begin{equation}
	|Z - z_{cv}|^2 = R^2 + \left|\frac{I_{cv}}{\gamma_1}\right|,
	\label{Eq:ZandRLneg}
\end{equation}
and we see that the center of vorticity must lie outside the circumcircle for all times.
In this case, there is a disk of radius $\sqrt{{I_{cv}}/{\gamma_1}}$ centerd at $z_{cv}$ 
that is a forbidden region for the circumcenter.

\section{Special solutions}\label{sec:special-sol}
The equations of motion for $R, A, B, C$ and $Z$ provide a different perspective on three-vortex motion than given by (\ref{Eq:dzdt}) or (\ref{Eq:Relmotion}).
We utilize this feature to extract some results on three-vortex motion that 
are more difficult to derive from other forms.

\subsection{Motions with constant $R$}
Motions for which $R$ is constant exist, e.g., the relative equilibria (and for the collinear relative equilibria $R$ is infinite).
One might ask if there are others. 
When $R$ is a constant, from \eqref{Eq:Intermediate} and \eqref{Eq:ds1dt} we have simplified expressions for $\dot A$, $\dot B$ and $\dot C$, {\it viz}
\begin{equation}
\begin{split}
&\frac{{\rm d}A}{{\rm d}t} = \frac{\Gamma_1}{8\pi R^2}\frac{\cot B-\cot C}{\cot A},\\
&\frac{{\rm d}B}{{\rm d}t} = \frac{\Gamma_2}{8\pi R^2}\frac{\cot C-\cot A}{\cot B},\\
&\frac{{\rm d}C}{{\rm d}t} = \frac{\Gamma_3}{8\pi R^2}\frac{\cot A-\cot B}{\cot C}.
\end{split}\label{Eq:ABCdotconstR}
\end{equation}

Now, the two dynamical integrals of these equations, (\ref{Eq:com}), take the reduced forms
\begin{subequations}
\begin{equation}
\frac{\sin^2A}{\Gamma_1} +\frac{\sin^2B}{\Gamma_2} +\frac{\sin^2C}{\Gamma_3} = \mbox{constant},
\end{equation}
and
\begin{equation}
\frac{\log\sin A}{\Gamma_1} + \frac{\log\sin B}{\Gamma_2}  + \frac{\log\sin C}{\Gamma_3} 
= -\frac{2\pi H}{\gamma_3}
= \lambda, 
\end{equation}
\end{subequations}
where $\lambda$ is a constant.
These integrals define surfaces in a $(\sin A, \sin B, \sin C)$-space, and 
these two surfaces intersect (at most) in a curve.
Adding to these equations the condition $A+B+C=\pi$ \eqref{Eq:interiorangles}, we see that we can at most expect isolated triples $(A,B,C)$ to give solutions, so that $\dot A= \dot B=\dot C=0$.
It then follows from \eqref{Eq:ABCdotconstR} 
that we must have $\cot A = \cot B = \cot C$.
This restriction yields the equilateral triangle, $A = B = C = \frac{\pi}{3}$.
The collinear relative equilibria are singular limits.
Thus, {\it there are no three-vortex motions with constant, finite $R$ other than the equilateral triangle configurations.}

We note that when $A=B=C=\pi/3$ (and for the collinear equilibria),  $\lambda=0$,
and thus $H=0$ is a
necessary condition for relative equilibria.


\subsection{Motions with invariant triangle shape}
By (\ref{Eq:Rdotfinal}), motions for which $A, B, C$ are constant imply either that $R^2$ is constant or that it grows linearly with time.
The former case yields the equilateral triangle relative equilibria.
In the latter case we have
\begin{equation}
R(t) = R_0 \sqrt{1-\frac{t}{\tau}},
\end{equation}
where $R_0$ is the initial value of $R$ and the time scale $\tau$ is given by
\begin{subequations}
\begin{equation}
\begin{split}
\frac{4\pi R^2_0}{\tau} =  -[&\Gamma_1\cot B\cot C(\cot B - \cot C)\ +\\
&\Gamma_2\cot C\cot A(\cot C - \cot A)\ +\\
&\Gamma_3\cot A\cot B(\cot A - \cot B)].
\end{split}
\end{equation}
But when $A, B, C$ are constants we have from (\ref{Eq:dAdt}) that the term in square brackets can be written in either of the forms
\begin{equation*}
\Gamma_1(\cot B-\cot C) =
\Gamma_2(\cot C-\cot A) = \Gamma_3(\cot A-\cot B).
\end{equation*}
Thus, we have
\begin{equation}
-\frac{4\pi R^2_0}{\tau} =  \Gamma_1(\cot B-\cot C) =
\Gamma_2(\cot C-\cot A) = \Gamma_3(\cot A-\cot B).
\end{equation}
These may be the simplest expressions known for $\tau$ which, when positive, gives the time of collapse~\cite{are2010}.
When $\tau<0$ we have self-similar expansion.
\end{subequations}

Equations (\ref{Eq:com}) enable us to read off the necessary and sufficient conditions for self-similar motion. The condition that
the angles $A, B, C$ be constant (but $R$ is not constant) means that $L=0$ and $\gamma_2=0$ in order for (\ref{Eq:com}) to be constants of
motion. On the other hand, if $L=0$ and $\gamma_2=0$, we get the following set of equations
\begin{subequations}
\begin{equation}
\frac{\sin^2A}{\Gamma_1} +\frac{\sin^2B}{\Gamma_2} +\frac{\sin^2C}{\Gamma_3} = 0
\end{equation}
\begin{equation}
\frac{\log\sin A}{\Gamma_1} + \frac{\log\sin B}{\Gamma_2}  + \frac{\log\sin C}{\Gamma_3} = \mbox{const.} = \lambda 
\label{Eq:com2H}
\end{equation}
\label{Eq:com2}
\end{subequations}
These two equations intersect in a curve, as we saw in the previous section. Together with the condition $A+B+C=\pi$ \eqref{Eq:interiorangles},
this means that there can at most be isolated triples as solutions. Thus, the angles $A, B, C$ must be constant. We conclude that 
$L=0$ and $\gamma_2=0$ are necessary and sufficient conditions for self-similar motion. 

We have seen in the previous section that $H=0$ is a necessary condition for equilibria. If we use $\lambda=0$ in (\ref{Eq:com2}), 
the equations become completely symmetrical in $A, B, C$. One solution to these pair of equations is the line, $\sin A=\sin B=\sin C$. Since these equations must meet in one
curve, we conclude that this is the only solution (which takes the form of a curve). One solution consistent with $A+B+C=\pi$
is then $A=B=C=\pi/3$. The other solutions are of course the collinear equilibria, with one of the angles being $\pi$ and the other angles 
being zero. Thus, we conclude that the conditions $H=0$, $L=0$ and $\gamma_2=0$ are necessary and sufficient conditions for relative equilibria of three vortices.


\section{Summary and outlook}

Starting from geometry and using the known equations of motion, we have shown that the motion of three vortices can be regarded as 
the motion of the center of vorticity with respect to a chosen origin of co-ordinates, plus a motion of the three vortices about the
center of vorticity. This latter motion is determined by the four autonomous equations (\ref{Eq:Rdotfinal}) and (\ref{Eq:dAdt}), 
along with the condition that the sum of the angles be equal to $\pi$. 
The transformation from the variables $z_\alpha$ to $Z$, $R$, and $\varphi_\alpha$
is not a canonical transformation. What this implies, however, is not clear. Also, the  number of variables in going from 
$Z$, $R$, and $\varphi_\alpha$ to $Z$, $R$, and $A, B, C$ is reduced by one. One of the implications of this reduction
is that rigid rotations of the relative equilibria configurations are lost in this representation.

We have examined some of the simple solutions to these equations
and have shown that these solutions exist under simple conditions that can be read off from the constants of motion. One might expect to 
find other solutions to these equations that throw light on the three vortex problem from a different perspective.




\begin{acknowledgements}
The authors dedicate this work to the memory of Hassan Aref, who passed away unexpectedly during the development 
of this manuscript, which was subsequently delayed for several years.  He is remembered fondly for his inspiration, insight, and friendship.
\end{acknowledgements}

\end{document}